\documentclass[10pt,journal,compsoc]{IEEEtran}
\ifCLASSOPTIONcompsoc
  \usepackage[nocompress]{cite}
\else
  \usepackage{cite}
\fi

\ifCLASSINFOpdf
\else
\fi
  \usepackage[caption=false,font=footnotesize]{subfig}
\usepackage{xcolor}
    \definecolor{webgreen}{rgb}{0,.5,0}
    \definecolor{webbrown}{rgb}{.6,0,0}
    \definecolor{webyellow}{rgb}{0.98,0.92,0.73}
    \definecolor{webgray}{rgb}{.753,.753,.753}
    \definecolor{webblue}{rgb}{0,0,.8}
    \definecolor{webred}{rgb}{0.8, 0, 0}   

\def\Arxiv{true}
\def\MayFloat{true}
\usepackage{pgfplots}
\usepackage{tikz}
\usetikzlibrary{arrows,shadows,backgrounds,calc,decorations.pathreplacing,decorations.pathmorphing,positioning, 
	automata}
\ifx\Arxiv\undefined	\usetikzlibrary{tikzmark} \fi
\usepackage{float}
\usepackage{colortbl}
\definecolor{ForestGreen}{rgb}{0.0, 0.4, 0.0}
\definecolor{webyellow}{rgb}{0.98,0.92,0.73}
\colorlet{HeadingColor}{ForestGreen}

\colorlet{tableheadcolor}{HeadingColor!25} 
\colorlet{tablerowcolor}{HeadingColor!10} 
\usepackage{url}
\usepackage{xkeyval}	
\usepackage{xcolor}
\definecolor{burntorange}{rgb}{0.8, 0.33, 0.0}
\definecolor{NavyBlue}{rgb}{0.0, 0.0, 0.5}
\definecolor{darkmagenta}{rgb}{0.55, 0.0, 0.55}
\definecolor{ivory}{rgb}{1.0, 1.0, 0.94}
\definecolor{ForestGreen}{rgb}{0.0, 0.4, 0.0}
\colorlet{SeparatorColor}{burntorange}

\usepackage{listingsutf8}
\usepackage{adjustbox}
\usepackage{booktabs}

\colorlet{HeadingColor}{ForestGreen}

	\makeatletter
	\define@key{MEMacros}{color}[green]{\def\ME@color{#1}}
	\define@key{MEMacros}{decorations}{\def\ME@decorations{#1}}
	\define@key{MEMacros}{iconheight}[10pt]{\def\ME@color{#1}}
	\define@key{MEMacros}{language}{\def\ME@language{#1}}
	\define@key{MEMacros}{number}{\def\ME@number{#1}}
	\define@key{MEMacros}{options}{\def\ME@options{#1}}
	\define@key{MEMacros}{plain}[true]{\def\ME@plain{#1}}		
	\define@key{MEMacros}{resize}[true]{\def\ME@resize{#1}}
	\define@key{MEMacros}{shrink}[true]{\def\ME@shrink{#1}}		
	\define@key{MEMacros}{tall}[true]{\def\ME@tall{#1}}
	\define@key{MEMacros}{title}{\def\ME@title{#1}}
	\define@boolkey{MEMacros}{wide}[true]{\def\ME@wide{#1}}
	\define@key{MEMacros}{width}[3cm]{\def\ME@width{#1}}
	\presetkeys{MEMacros}{color=green}{}%
	\presetkeys{MEMacros}{number=1}{}%
	\presetkeys{MEMacros}{plain=false}{}%
	\presetkeys{MEMacros}{resize=false}{}%
	\presetkeys{MEMacros}{shrink=false}{}%
	\presetkeys{MEMacros}{tall=false}{}%
	\presetkeys{MEMacros}{wide=false}{}%
	\presetkeys{MEMacros}{width=3cm}{}%
\makeatother

\ifx\MayFloat\undefined
\else
	\newfloat{program}{thp}{lpp}[section]
	\floatstyle{plaintop}
\fi

\newlength{\mywidth} 

\makeatletter

\newlength{\FigWidth}
\makeatletter
\newcommand\MESourceFile[4][]{
	\setkeys{MEMacros}{wide=false,language={[ANSI]C},options={}, decorations={},#1}%
	\MESetStandardListingFormat
	\ifx\MayFloat\undefined 
		\def\startsource{
			\setlength{\FigWidth}{\textwidth}
		{\color{HeadingColor}
			}
			\par
		}
		\def\stopsource{}
	\else 
		\ifKV@MEMacros@wide 
		   \def\startsource{
			\setlength{\FigWidth}{\textwidth} 
			\begin{program*}[!hbt] 
			}
			\def\stopsource{\end{program*}}			
		\else 
			\if@twocolumn
			  \setlength{\FigWidth}{\columnwidth}
			\else
			  \setlength{\FigWidth}{.7\textwidth}
			 \fi
			\def\startsource{\begin{program}[!hbt]}
			\def\stopsource{\end{program}}
		\fi
	\fi
	\noindent\startsource\hskip-1em
	\vspace{-10pt}
	{
		\noindent\begingroup\protected@edef\x{\endgroup\noexpand
			\lstinputlisting[language={\ME@language}, \ME@options, label=#4, name=#4,
			caption ={#3}
			]{#2}}
		\x
		\ME@decorations 
	}					
	\stopsource
}

\makeatletter  
\newcommand\MEtable[6][]{%
	\setkeys{MEMacros}{wide=false,#1}%
	\ifx\MayFloat\undefined
		\def\startsource{
			{\color{HeadingColor}\bfseries\scriptsize #3}\par\vskip-1.8\baselineskip}
		\setlength{\mywidth}{#6\textwidth}
		\def\stopsource{}
	\else
		\ifKV@MEMacros@wide
			\def\startsource{\begin{table*}[!hbtp]	}
			\def\stopsource{\end{table*}}
			\setlength{\mywidth}{#6\textwidth}
		\else
			\setlength{\mywidth}{#6\columnwidth}
			\ifx\WEBBook\undefined\else\setlength{\mywidth}{.5\mywidth}\fi
			\ifx\eBook\undefined\else\setlength{\mywidth}{.7\mywidth}\fi
			\def\startsource{\begin{table}[!hbtp]}
			\def\stopsource{\end{table}}
		\fi
	\fi
	\startsource	  
		\begin{center}
			\vbox{
				\ifx\WEBBook\undefined\else\vskip-1.3\baselineskip\fi
				\begin{minipage}{\mywidth} 			
					\ifx\MayFloat\undefined
					\else{\color{HeadingColor}\caption{#3}\label{#4}}
					\fi 
				\end{minipage}
				\vglue-0.2\baselineskip\par
				\begin{minipage}{\mywidth}
					\ifthenelse{\equal{#5}{}}{}{\vskip-.1\baselineskip{\tiny \copyright #5}	\hfill
						}  
				\end{minipage}
				\par\vskip-.7\baselineskip
				\noindent\makebox[\mywidth]{\color{SeparatorColor}\rule{\mywidth}{1pt}}\par
						\maxsizebox{\mywidth}{.75\textheight}{ 
						#2}
				\par\noindent\makebox[\mywidth]{\color{SeparatorColor}\rule{\mywidth}{1pt}}\par
				\par\vskip-.7\baselineskip
			}
		\end{center}
	\ifx\LecturePrintable\undefined\else\par\vskip-.8\baselineskip\fi
	\stopsource%
}

\makeatother

\ifx\eBook\undefined
	\def\lstsize{\scriptsize}
\else
	\def\lstsize{\tiny}
\fi

\newcommand{\MESetStandardListingFormat}
{
	\lstset{
	    literate=
         {í}{{\'i}}1
         {Í}{{\'I}}1
	    {ő}{{\H{}o}}1
	    {Ő}{{\H{}O}}1
	    {Ű}{{\H{}U}}1
	    {ű}{{\H{}u}}1
	    	}	    
	\lstset{
		numberstyle=\tiny,          
		numbersep=5pt,              
		tabsize=2,                 	 
		inputencoding=utf8/latin2,	
		extendedchars=true,         %
		escapechar=\@,
		breaklines=true,        
		columns=fullflexible,  
		breakatwhitespace=true,    
        keepspaces=true,
        escapeinside={\%*}{*)},          
		frame=tb, 
		framerule=.5pt, 
		rulecolor= \color{SeparatorColor},
		backgroundcolor=\color{ivory},
		basicstyle=\ttfamily\color{black}\normalsize\bfseries, 
		keywordstyle=\bfseries\color{darkmagenta},
		identifierstyle=\bfseries\color{NavyBlue},
		commentstyle=\itshape\bfseries\color{ForestGreen},
		stringstyle=\itshape\bfseries\color{burntorange}, 
		lineskip=0pt,aboveskip=4pt,belowskip=2pt,
		framesep=4pt,rulesep=2pt, 
		showspaces=false,           %
		showtabs=false,             %
		framexleftmargin=0pt,
		framexrightmargin=0pt,
		showstringspaces=false      
	}	
}
\newcommand\MESetListingFormat[2][]	
{
    \MESetStandardListingFormat
    \lstset
    	{
		#1,
		language={#2},		
         numberstyle=\tiny,          
         numbersep=5pt,              
         tabsize=2,                 	 
         inputencoding=utf8/latin2,	
         extendedchars=true,         %
         escapechar=\@,
		breaklines=true,        
		breakatwhitespace=true,    
		escapeinside={\%*}{*)},          
		frame=tb, 
		framerule=.5pt, 
		rulecolor= \color{SeparatorColor},
		 backgroundcolor=\color{ivory},
       basicstyle=\ttfamily\color{black}\lstsize , 
         keywordstyle=\bfseries\color{magenta},
         identifierstyle=\bfseries\color{NavyBlue},
 		commentstyle=\itshape\bfseries\color{ForestGreen},
        stringstyle=\itshape\bfseries\color{burntorange}, 
          lineskip=0pt,aboveskip=4pt,belowskip=2pt,
            framesep=4pt,rulesep=2pt, 
          showspaces=false,           %
         showtabs=false,             %
         xleftmargin=-1pt,
         framexbottommargin=4pt,
         framextopmargin=4pt,
         gobble=5,
		framexleftmargin=0pt,
		framexrightmargin=0pt,
         showstringspaces=false      
	}	
}


\makeatletter
\newcounter{qan}\newcounter{qano}
\newcommand\MEListBalls[3][]{%
\ifx\Arxiv\undefined 
	\setkeys{MEMacros}{color=orange,#1}%
	\setkeys{MEMacros}{number=1,#1}%
\setcounter{qan}{0}
\setcounter{qano}{0}
   \begingroup%
    \foreach \x in {#3}
    {  \addtocounter{qan}{1}
    	\addtocounter{qano}{1}
  {\tikz[remember picture,overlay]
    {\expandafter\node[circle, inner sep=2pt, draw,fill=\ME@color,ball color=\ME@color, shading=ball, font=\scriptsize\bfseries, drop shadow]
   at  ([xshift=+10pt,yshift=+2pt]{pic cs:line-#2-\x-end}) {\lstsize\arabic{qan}};\expandafter}}
   }
	\endgroup
\fi
}
\makeatother

\lstdefinelanguage
   [x64]{Assembler}     
   [x86masm]{Assembler} 
   {morekeywords={CDQE,CQO,CMPSQ,CMPXCHG16B,JRCXZ,LODSQ,MOVSXD, %
                  POPFQ,PUSHFQ,SCASQ,STOSQ,IRETQ,RDTSCP,SWAPGS, %
                  rax,rdx,rcx,rbx,rsi,rdi,rsp,rbp, %
                  r8,r8d,r8w,r8b,r9,r9d,r9w,r9b}} 

\lstdefinelanguage
   [y86]{Assembler}     
   [x86masm]{Assembler} 
   {morekeywords={ halt, nop, movl, rrmovl, irmovl, rmmovl, mrmovl, rrmovl, jump,%
                  xorl, andl, je, addl, jne, subl, pushl, popl, jXX,%
                  cmovXX, cmovl, cmovle, cmove, cmovne, cmovge, cmovg, return,%
                   rax,rbx,rcx,rdx,xmm0,rbp, load,mulss,%
                   OPl, ZF, SF, OF,%
                   eno, ecc, esv,
                   QCreate, QTCreate, QFCreate, QTerm, QCall, QAlloc,
                   QWait, QPWait, QIWait, QInt
                  },
      morecomment=[l]{\#},
      sensitive=false,
     } 

\newcommand{\MEBall}[2]{%
{\tiny#1}\hskip-4pt~\raisebox{-.08cm}{
\tikz \node (1ex,1ex)
 [
circle,draw,ball color=green, shading=ball,  font=\bfseries, scale=0.55] {#2};}%
~\hskip-4pt}

		\usepackage[toc,acronym]{glossaries}

		\makeglossaries

  \newacronym{ALU}{ALU}{Arithmetic and Logic Unit}
  \newacronym{CMP}{CMP}{Chip Multi Processor}
  \newacronym{CPU}{CPU}{Central Processing Unit}
  \newacronym{CSTB}{CSTB}{Computer Science and Telecommunications Board}
  \newacronym{EDVAC}{EDVAC}{Electronic Discrete Variable Computer}
  \newacronym{ENIAC}{ENIAC}{Electronic Numerical Integrator and Computer}
  \newacronym{EMPA}{EMPA}{Explicitly Many-Processor Approach}
  \newacronym{GPGPU}{GPGPU}{General-Purpose Graphics Processing Unit}
  \newacronym{HW}{HW}{hardware   }
  \newacronym[description={Instruction Level Parallelism}]{ILP}{ILP}{Instruction Level Parallelism}
  \newacronym{I/O}{I/O}{input/output}
  \newacronym[description={Instruction Set Architecture}]{ISA}{ISA}{Instruction Set Architecture}
\newacronym[description={Million Instructions Per Second}]
{MIPS}{MIPS}{Million Instructions Per Second}
 \newacronym[description={Object Oriented Programming}]{OOP}{OOP}
 {Object Oriented Programming}

  \newacronym[description={Operating System}]{OS}{OS}{Operating System}
  \newacronym[description={Program Counter}]{PC}{PC}{Program Counter}
  \newacronym{PRAM}{PRAM}{Parallel Random Access Model \protect{\cite{Vishkin:FineGrainedProgramming} }
  }
  \newacronym{PU}{PU}{Processing Unit}
  \newacronym{QT}{QT}{Quasi-Thread}
  \newacronym{RC}{RC}{reconfigurable}
  \newacronym{RT}{RT}{real time}
  \newacronym{SIMD}{SIMD}{Single Instruction Multiple Data}
  \newacronym{SPA}{SPA}{Single Processor Approach}
  \newacronym{SV}{SV}{supervisor}
  \newacronym{SW}{SW}{software}
  \newacronym[description={Thread Level Parallelism}]{TLP}{TLP}{Thread Level Parallelism}

\newacronym{XMT}{XMT}{eXplicit Multi-Threading}

\usepackage{listings}

\makeatletter
\newcommand{\highlighted}[2][]{%
	\setkeys{MEMacros}{color=darkmagenta!40!black!80,#1}%
	{\textit{\textbf
			{\textcolor{\expandafter\ME@color\expandafter}{#2}}}}%
}

\newcommand\MEfigure[6][]{
	\setkeys{MEMacros}{wide=false,#1}%
	\ifx\MayFloat\undefined
		\def\startsource{ 
		}
		\setlength{\mywidth}{#6\textwidth}
		\def\stopsource{}
	\else
		\ifKV@MEMacros@wide
			\def\startsource{\par\begin{figure*}[!hbt]	
					}
			\def\stopsource{\end{figure*}}
			\setlength{\mywidth}{#6\textwidth}
		\else
			\setlength{\mywidth}{#6\columnwidth}
			\ifx\WEBBook\undefined\else\setlength{\mywidth}{.55\mywidth}\fi
			\ifx\eBook\undefined\else\setlength{\mywidth}{.7\mywidth}\fi
			\def\startsource{\par\begin{figure}[!hbtp]}
			\def\stopsource{\end{figure}}
		\fi
	\fi
	
	\par\startsource	  
	{{\centering
	
			\vbox{
				\ifx\WEBBook\undefined\else\vskip-.3\baselineskip\fi
				\begin{minipage}{\mywidth}
					\ifthenelse{\equal{#5}{}}{}{\vskip-.1\baselineskip{\tiny \copyright #5}	\hfill
						}  
				\end{minipage}
				\par\vskip-.9\baselineskip
				\noindent\makebox[\mywidth]{\color{SeparatorColor}\rule{\mywidth}{1pt}}\par\vskip.2\baselineskip
						\maxsizebox{\mywidth}{.75\textheight}
						{ 

						\includegraphics[width=\mywidth,keepaspectratio]{#2}
						}
					\par\vskip-.8\baselineskip\par
				\noindent\makebox[\mywidth]{\color{SeparatorColor}\rule{\mywidth}{1pt}}\par
			
				\begin{minipage}{\mywidth} 			
					\ifx\MayFloat\undefined
						{\scriptsize {\color{HeadingColor}#3}}
					\else{\color{HeadingColor}\caption{{#3}}\label{#4}}
					\fi 
				\end{minipage}
			}
		
	\ifx\LecturePrintable\undefined\else\par\vspace{-15pt}\fi				
}}
	\stopsource%
}

\newcommand\MEtikzfigure[6][]{
	\setkeys{MEMacros}{wide=false,#1}%
	\ifx\MayFloat\undefined
		\def\startsource{ 
		}
		\setlength{\mywidth}{#6\textwidth}
		\def\stopsource{}
	\else
		\ifKV@MEMacros@wide
			\def\startsource{\par\begin{figure*}[!hbt]	
					}
			\def\stopsource{\end{figure*}}
			\setlength{\mywidth}{#6\textwidth}
		\else
			\setlength{\mywidth}{#6\columnwidth}
			\ifx\WEBBook\undefined\else\setlength{\mywidth}{.55\mywidth}\fi
			\ifx\eBook\undefined\else\setlength{\mywidth}{.7\mywidth}\fi
			\def\startsource{\par\begin{figure}[!hbtp]}
			\def\stopsource{\end{figure}}
		\fi
	\fi
	
	\par\startsource	  
	{{\centering
	
			\vbox{
				\ifx\WEBBook\undefined\else\vskip-.3\baselineskip\fi
				\begin{minipage}{\mywidth}
					\ifthenelse{\equal{#5}{}}{}{\vskip-.1\baselineskip{\tiny \copyright #5}	\hfill
						}  
				\end{minipage}
				\par\vskip-.9\baselineskip
				\noindent\makebox[\mywidth]{\color{SeparatorColor}\rule{\mywidth}{1pt}}\par\vskip.2\baselineskip
						\maxsizebox{\mywidth}{.75\textheight}
						{ 
						#2
						}
				\noindent\makebox[\mywidth]{\color{SeparatorColor}\rule{\mywidth}{1pt}}\par
			
				\begin{minipage}{\mywidth} 			
					\ifx\MayFloat\undefined
						{\scriptsize {\color{HeadingColor}#3}}
					\else{\color{HeadingColor}\caption{{#3}}\label{#4}}
					\fi 
				\end{minipage}
			}
		
}}
	\stopsource%
}

\makeatother 
\makeatletter
\newcommand{\gettikzxy}[3]{%
  \tikz@scan@one@point\pgfutil@firstofone#1\relax
  \edef#2{\the\pgf@x}%
  \edef#3{\the\pgf@y}%
}
\makeatother

\begin{document}
\title{A new kind of parallelism and its programming\\in the Explicitly Many-Processor Approach}
\author{J\'anos V\'egh
\IEEEcompsocitemizethanks{\IEEEcompsocthanksitem J. V\'egh is with Faculty of Mechanical Engineering and Informatics, University of Miskolc, Hungary\protect\\
E-mail: J.Vegh@uni-miskolc.hu
}
\thanks{Manuscript received July 10, 2016; revised August 26, 2016.}}

\markboth{Programming for EMPA,~Vol.~14, No.~8, August~2016}%
{Shell \MakeLowercase{\textit{et al.}}: Bare Demo of IEEEtran.cls for Computer Society Journals}
%

\IEEEtitleabstractindextext{%
\begin{abstract}
The processor accelerators are effective because they are  working
not (completely) on principles of stored program computers.
They use some kind of parallelism, and it is rather hard
to program them effectively: a parallel architecture
by means of (and thinking in) sequential programming.
The recently introduced \gls{EMPA} architecture uses a new kind of parallelism, which offers
the potential of reaching higher degree of parallelism,
and also provides
extra possibilities and challenges. It not only provides synchronization and inherent parallelization,
but also takes over some duties typically offered by the \gls{OS}, and even opens
the till now closed machine instructions for the end-user.
A toolchain for \gls{EMPA} architecture with Y86 cores has been prepared,
including an assembler and a cycle-accurate simulator.
The assembler is equipped with some meta-instructions, which allow to use
all advanced possibilities of the \gls{EMPA} architecture, and at the 
same time provide a (nearly) conventional-style programming.
The cycle accurate simulator is able to execute the \gls{EMPA}-aware object code,
and is a good tool for developing algorithms for \gls{EMPA}.

\end{abstract}

}

\maketitle
\lstset{language={[x86masm]Assembler}, basicstyle=\ttfamily\color{black}\normalsize}

\IEEEdisplaynontitleabstractindextext

\IEEEpeerreviewmaketitle

\IEEEraisesectionheading{\section{Introduction}\label{sec:introduction}}

	\MESetStandardListingFormat

%
%
%
%
\IEEEPARstart{P}{rogramming} hardware accelerators is a real challenge.
The accelerator is always outside the processor, and it is efficient because
it does not work as the conventional, programmable processors do. 
Several problems must be solved in order to connect two different worlds:
the stored program (von Neumann) processors with rest of the world.
It is a challenge for the \gls{HW} designers: the principle of operation 
of the processing units based on the von Neumann principles has
inherent inefficiencies~\cite{VeghEMPA:2016}, and also using those external accelerators
from software running on a conventional architecture is only possible
in rather time-consuming ways~\cite{hallaron}.
It is a challenge also for the \gls{SW} designers: the program languages are structured
according to the von Neumann principles~\cite{BackusNeumannProgrammingStyle},
and the programs need to handle facilities, different from the ones, 
they were designed for.
Even the linking method is hard to select. For universal utilization
(and also because of the compactness of the \gls{CPU}s), the accelerators 
are usually implemented as \gls{I/O} devices. However, since the \gls{OS}s
must provide protection for the \gls{I/O} operations, which is a time-consuming 
procedure~\cite{hallaron}, only longer code fragments
can be delegated to the accelerators.

The modern many-core systems could serve as good starting point to develop
general purpose accelerators, using new forms of parallelization, but
mainly their programming provokes technical~\cite{Larus:ProgrammingMulticoreCACM},
efficiency~\cite{HWcontrolledthreadsMahesri:2007}
and performance~\cite{NinjaPerformanceGap:2015:CACM} questions;
so that trend seems to be broken~\cite{Viskin:ViewpointProgrammingMulticore}.
The new kind of parallelism, introduced below, allows to approach
the theoretically possible maximal parallelism and at the same time to
simplify the \gls{HW} construction, but of course 
requires non-conventional processor architecture.

The EMPA architecture~\cite{VeghEMPA:2016} seems to be especially hard to program:
the processor architecture can be configured by the end-user, and so the architecture
may change continuously during the operation; the cores
communicate with each other; synchronized internal data transfer between cores takes place;
different program parts run in parallel, which must handle data and control dependencies;
there are many program counters, belonging to independently running cores;
even the conventionally closed unit "machine instruction" can be opened for other
processing units, and so on.
For the first look, it does not seem possible to program it using facilities
mostly similar to the conventional programming. The paper introduces a new kind of parallelism, as well as a programming
language and methodology, which allows to utilize the enhanced performance of
the \gls{EMPA} processors, while the programming methods remain as close as possible
to the conventional ones.

\section{The dynamic parallelism}

The parallelism assumes the presence of several processing units, and the 
reachable speedup of course strongly depends on the availability of the 
corresponding \gls{HW} units. 
Let us suppose we want to calculate expressions (see \cite{HwangParallelism:2016})
\par\noindent\vskip-\parskip\lstinline|A = (C*D)+(E*F)|
\par\noindent\vskip-\parskip\lstinline|B = (C*D)-(E*F)|
\par\noindent where we have altogether 4 load operations, 2 multiplications,
and 2 additions. 

\newlength{\nd}	
\setlength{\nd}{1.5cm}

\MEtikzfigure[]{
\begin{tabular}{cc}
\begin{tikzpicture}[->,>=stealth',shorten >=1pt,auto,node distance=\nd and \nd,
                    semithick]
  \tikzstyle{every state}=[scale=0.75]

  \node[state]  (L1)       {$L_1$};
  \node         (E1) [ above of=L1]      {$B = (C*D)-(E*F)$};
  \node         (E2m) [ above of=E1]      {};
  \node         (E2) at ($(E1)!0.5!(E2m)$)     {$A = (C*D)+(E*F)$};
  
  \node[state]  (L2) [ right of=L1]     {$L_2$};
  \node[state]  (L3) [ right of=L2]     {$L_3$};
  \node[state]  (L4) [ right of=L3]     {$L_4$};

  \node  (X1F)  at ($(L1)!0.5!(L2)$)     {};
  \node[state]  (X1)   [ below of=X1F]     {$X_1$};
  \node  (X2F)  at ($(L3)!0.5!(L4)$)     {};
  \node[state]  (X2)   [ below of=X2F]     {$X_2$};
 
  \node[state]  (plus) [ below of=X1]     {$+$};
  \node[state]  (minus) [ below of=X2]     {$-$};
  
  \node (A) [ below of=plus]     {$A$};
  \node (B) [ below of=minus]     {$B$};

 \path 
   (L1) edge  (X1)
   (L2) edge  (X1)
   (L3) edge  (X2)
   (L4) edge  (X2);
 \path 
   (X1) edge  (plus)
   (X1) edge  (minus)
   (X2) edge  (plus)
   (X2) edge  (minus);
 \path 
   (plus) edge  (A)
   (minus) edge  (B);

  \node         (C1) [ left of=L1] {$Cycle\ 1$};
\gettikzxy{(C1)}{\cx}{\cy}
\gettikzxy{(X1)}{\xx}{\xy}
\gettikzxy{(plus)}{\px}{\py}

  \node         (C2) at ($(\cx,\xy)$) {$Cycle\ 2$};
  \node         (C3) at ($(\cx,\py)$) {$Cycle\ 3$};

\end{tikzpicture}
&
\begin{tikzpicture}[->,>=stealth',shorten >=1pt,auto,
node  distance=\nd and \nd,
                    semithick]
  \tikzstyle{every state}=[scale=0.75]
  \node[state]  (L1)                    {$L_1$};
  \node[state]  (L2) [ below of=L1]     {$L_2$};
  \node[state]  (L3) [ below of=L2]     {$L_3$};
  \node[state]  (L4) [ below of=L3]     {$L_4$};

  \node[state]  (X1) [ left of=L3]     {$X_1$};
\phantom{  \node[state]        (XM) [ below of=X1]     {$X_m$};}
  \node[state]  (X2) [ below of=XM]     {$X_2$};

  \node[state]  (plus) [ below of=X2]     {$+$};
  \node[state]  (minus) [ below of=plus]     {$-$};
  \node         (B)    [ below of=minus]     {$B$};
  \node         (Bb)    [ left of=minus]     {};
  \node         (A)    [ right of=B]    {$A$};

  \node         (C1) [ right of=L1] {$Cycle\ 1$};
\gettikzxy{(C1)}{\cxa}{\cya}
\gettikzxy{(L2)}{\lxa}{\lya}
  \node         (C2) at ($(\cxa,\lya)$)  {$Cycle\ 2$};
\gettikzxy{(L3)}{\lxb}{\lyb}
  \node         (C3) at ($(\cxa,\lyb)$) {$Cycle\ 3$};
\gettikzxy{(L4)}{\lxc}{\lyc}
  \node         (C4) at ($(\cxa,\lyc)$) {$Cycle\ 4$};
\gettikzxy{(X2)}{\xx}{\xy}
  \node         (C5) at ($(\cxa,\xy)$) {$Cycle\ 5$};
\gettikzxy{(plus)}{\px}{\py}
  \node         (C6) at ($(\cxa,\py)$) {$Cycle\ 6$};
\gettikzxy{(minus)}{\mx}{\my}
  \node         (C6) at ($(\cxa,\my)$) {$Cycle\ 7$};

\gettikzxy{(X1)}{\xx}{\xy}
\gettikzxy{(plus)}{\px}{\py}
 \path 
   (L1) edge  (X1)
   (L2) edge  (X1)
   (X2) edge (plus);
 \path 
   (L3) edge  (X2)
   (L4) edge  (X2);
 \path 
   (X1) edge [bend right] node {} (plus)
   (X1) edge [bend right] node {} (minus);
 \path 
   (X2) edge [bend left] node {} (minus);

 \path 
   (plus) edge  (A)
   (minus) edge  (B);

\end{tikzpicture}
\\
\end{tabular}
	}
{The theoretical parallelism (left) vs parallelism implemented on a two-issue fixed architecture processor (right)}
{fig:SWvsHWparallelism}{}{}

\subsection{Theoretical (software) parallelism}

The theoretical (or \gls{SW}) parallelism 
only considers the different kinds of dependences between the values and operations
(control and data dependences), and assumes the presence of the needed number of \gls{HW} units;
i.e. it provides a kind of upper bound for the reachable parallelism.
For calculating the theoretical parallelism, one can assume that
we have a processor, which has (at least) 4 memory access units, 2 multipliers
and 2 adders (or equally: an at least 4-issue universal processor). With such a processor (see Fig.~\ref{fig:SWvsHWparallelism}) we can
load all the four operands in the first machine cycle, to
do the multiplications in the second cycle, and to make the addition/subtraction in the last cycle.
\index{parallelism!theoretical}

\subsection{Multiple-issue processor parallelism}

The real processors, however, are not built with arbitrarily large  number of processing units.
In practice, the processors may be built with having so called multiple-issue, single pipeline architecture,
i.e. in the same cycle can execute more than one operations, if there are available 
processing units  which are able to perform the requested operation. 
A so called two-issue processor can make (for example) an arithmetic and a memory access
operation at once.

Before making the first multiplication (see Fig~\ref{fig:SWvsHWparallelism}, right side), in the first two cycles the processor can load
the two operands, and in the third cycle, it can make the first multiplication. 
During the multiplication, the memory access unit is free, so it can load
simultaneously the third operand. In the fourth cycle, the fourth operand is loaded,
and so finally the second operand for the second multiplication is provided (the first operand is waiting
since the third cycle).
In the fifth cycle the second multiplication can be performed, and so for the sixth cycle result $A$
is provided, and similarly for the seventh cycle result $B$ is also available. Notice that both the memory
access and the aritmetic units are only utilized in 4 cycles (out of the 7), in 3 machine cycles they are unused. Only cycle 3 is when both units are in use.
\index{parallelism!muliple-issue}

\subsection{Dual core parallelism}

\MEtikzfigure
{
\begin{tabular}{cc}
\begin{tikzpicture}[->,>=stealth',shorten >=1pt,auto,node distance=\nd and \nd,
                    semithick]
  \tikzstyle{every state}=[scale=0.75]

  \node[state]  (L1)       {$L_1$};
  \node         (E1) [ above of=L1]      {$B = (C*D)-(E*F)$};
  \node         (E2m) [ above of=E1]      {};
  \node         (E2) at ($(E1)!0.5!(E2m)$)     {$A = (C*D)+(E*F)$};

  \node[state]  (L2) [ right of=L1]     {$L_2$};
  \node[state]  (L3) [ right of=L2]     {$L_3$};
  \node[state]  (L4) [ right of=L3]     {$L_4$};

  \node  (X1F)  at ($(L1)!0.5!(L2)$)     {};
  \node[state]  (X1)   [ below of=X1F]     {$X_1$};
  \node  (X2F)  at ($(L3)!0.5!(L4)$)     {};
  \node[state]  (X2)   [ below of=X2F]     {$X_2$};
 
  \node[state]  (plus) [ below of=X1]     {$+$};
  \node[state]  (minus) [ below of=X2]     {$-$};
  
  \node (A) [ below of=plus]     {$A$};
  \node (B) [ below of=minus]     {$B$};

 \path 
   (L1) edge  (X1)
   (L2) edge  (X1)
   (L3) edge  (X2)
   (L4) edge  (X2);
 \path 
   (X1) edge  (plus)
   (X1) edge  (minus)
   (X2) edge  (plus)
   (X2) edge  (minus);
 \path 
   (plus) edge  (A)
   (minus) edge  (B);

  \node         (C1) [ left of=L1] {$Cycle\ 1$};
\gettikzxy{(C1)}{\cx}{\cy}
\gettikzxy{(X1)}{\xx}{\xy}
\gettikzxy{(plus)}{\px}{\py}

  \node         (C2) at ($(\cx,\xy)$) {$Cycle\ 2$};
  \node         (C3) at ($(\cx,\py)$) {$Cycle\ 3$};

\end{tikzpicture}
&
\begin{tikzpicture}[->,>=stealth',shorten >=1pt,auto,node distance=\nd and\nd,
                    semithick]
  \tikzstyle{every state}=[scale=0.75]

  \node[state]  (L1)                    {$L_1$};
  \node         (L1F)[right of=L1]     {};
  \node[state]  (L3) [ right of=L1F]     {$L_3$};
  \node[state]  (L2) [ below of=L1]     {$L_2$};
  \node[state]  (L4) [ below of=L3]     {$L_4$};

  \node[state]  (X1)   [ below of=L2]     {$X_1$};
  \node[state]  (X2)   [ below of=L4]     {$X_2$};

 \path 
   (L1) edge [bend right] node {} (X1)
   (L3) edge [bend left] node {} (X2)
   (L2) edge             node {} (X1)
   (L4) edge             node {} (X2);

  \node[state,fill=webyellow]  (S1)   [ below of=X1]     {$S_1$};
  \node[state,fill=webyellow]  (S2)   [ below of=X2]     {$S_2$};
 \path 
   (X1) edge             node {} (S1)
   (X2) edge             node {} (S2);
  
  \node[state,fill=webyellow]  (L5)   [ below of=S1]     {$L_5$};
  \node[state,fill=webyellow]  (L6)   [ below of=S2]     {$L_6$};
 \path 
   (S1) edge             node {} (L6)
   (S2) edge             node {} (L5);

  \node[state]  (plus1)   [ below of=L5]     {$+$};
  \node[state]  (minus)   [ below of=L6]     {$-$};
 \path 
   (S1) edge [bend right] node {} (plus1)
   (S2) edge [bend left] node {} (minus);

  \node  (A)   [ below of=plus1]     {$A$};
  \node  (B)   [ below of=minus]     {$B$};
 \path 
   (plus1) edge             node {} (A)
   (minus) edge             node {} (B);
 \path 
   (L5) edge             node {} (plus1)
   (L6) edge             node {} (minus);

  \node  (X1F)  at ($(L1)!0.5!(L3)$)     {};
  \node  (X2F)  at ($(A)!0.5!(B)$)     {};
 \path[dashed]
   (X1F) edge node {} (X2F);

  \node         (C1) [ right of=L3] {$Cycle\ 1$};
\gettikzxy{(C1)}{\cxa}{\cya}
\gettikzxy{(L4)}{\lxa}{\lya}
  \node         (C2) at ($(\cxa,\lya)$)  {$Cycle\ 2$};
\gettikzxy{(X2)}{\lxb}{\lyb}
  \node         (C3) at ($(\cxa,\lyb)$) {$Cycle\ 3$};
\gettikzxy{(S2)}{\lsb}{\lsb}
  \node         (C4) at ($(\cxa,\lsb)$) {$Cycle\ 4$};
\gettikzxy{(L6)}{\lxc}{\lyc}
  \node         (C5) at ($(\cxa,\lyc)$) {$Cycle\ 5$};
\gettikzxy{(minus)}{\mx}{\my}
  \node         (C5) at ($(\cxa,\my)$) {$Cycle\ 6$};

\end{tikzpicture}
\\
\end{tabular}
}
{The theoretical parallelism (left) vs parallelism implemented on a single-issue dual processor system (right).
}
{fig:dualproc}{}{}

One might think that using two independent, single issue processors communicating through
shared memories can be equally good 
for solving the sample task. Initially, both processors
can load their arguments (see Fig.~\ref{fig:dualproc})  and make their multiplication.
However, after those operations processors must share their result
with their party, i.e. they 
store their result in the shared memory,
and load the result stored by their party from the shared memory.
For doing so, store ($S_i$) and load ($L_i$) operations must be inserted 
in the chain of operations of both processors. These (essentially obsolete, but needed for the communication) operations increase the number of
operations to 12, and so both processors must execute 6 cycles. Compare it to the 
7 cycles of a 2-issue single-processor system above. Obviously, investing into the 
second processor and shared memory \gls{HW}, does not result in the expected increase
of performance (in addition, the memory access operations are very expensive
in terms of execution time; and also we can only hope that the operand
the processor reads was already written by the other party).
\index{parallelism!dual core}

\MEtikzfigure[wide]
{
\begin{tabular}{cc}
\begin{tikzpicture}[->,>=stealth',shorten >=1pt,auto,node distance=\nd and \nd,
                    semithick]
  \tikzstyle{every state}=[scale=0.75]
  \node[state]  (L1)       {$L_1$};
  \node         (E1) [ above of=L1]      {$B = (C*D)-(E*F)$};
  \node         (E2m) [ above of=E1]      {};
  \node         (E2) at ($(E1)!0.5!(E2m)$)     {$A = (C*D)+(E*F)$};

  \node[state]  (L2) [ right of=L1]     {$L_2$};
  \node[state]  (L3) [ right of=L2]     {$L_3$};
  \node[state]  (L4) [ right of=L3]     {$L_4$};

  \node  (X1F)  at ($(L1)!0.5!(L2)$)     {};
  \node[state]  (X1)   [ below of=X1F]     {$X_1$};
  \node  (X2F)  at ($(L3)!0.5!(L4)$)     {};
  \node[state]  (X2)   [ below of=X2F]     {$X_2$};
 
  \node[state]  (plus) [ below of=X1]     {$+$};
  \node[state]  (minus) [ below of=X2]     {$-$};
  
  \node (A) [ below of=plus]     {$A$};
  \node (B) [ below of=minus]     {$B$};

 \path 
   (L1) edge  (X1)
   (L2) edge  (X1)
   (L3) edge  (X2)
   (L4) edge  (X2);
 \path 
   (X1) edge  (plus)
   (X1) edge  (minus)
   (X2) edge  (plus)
   (X2) edge  (minus);
 \path 
   (plus) edge  (A)
   (minus) edge  (B);

  \node         (C1) [ left of=L1] {$Cycle\ 1$};
\gettikzxy{(C1)}{\cx}{\cy}
\gettikzxy{(X1)}{\xx}{\xy}
\gettikzxy{(plus)}{\px}{\py}

  \node         (C2) at ($(\cx,\xy)$) {$Cycle\ 2$};
  \node         (C3) at ($(\cx,\py)$) {$Cycle\ 3$};

\end{tikzpicture}
&
\begin{tikzpicture}[->,>=stealth',shorten >=1pt,auto,
node  distance=\nd and \nd,
                    semithick]
  \tikzstyle{every state}=[scale=0.75]
  
  \node[state,fill=webyellow]  (O1)                    {$O_1$};
  \node[state,fill=webyellow]  (H1) at ($(O1)+(\nd,-.2\nd)$)       {$H_1$};
  \node[state,fill=webyellow]  (H2) at ($(O1)+(4\nd,-.4\nd)$)       {$H_2$};
   \path 
     (O1) edge [bend left] node {} (H1)
     (O1) edge [bend left] node {} (H2);
  
  \node[state]  (L11) at ($(H1)+(\nd,-.2\nd)$)       {$L_{11}$};
  \node[state]  (L12) at ($(H1)+(2\nd,-.4\nd)$)       {$L_{12}$};
   \path 
     (H1) edge [bend left] node {} (L11)
     (H1) edge [bend left] node {} (L12);

  \node[state]  (L21) at ($(H2)+(\nd,-.2\nd)$)       {$L_{21}$};
  \node[state]  (L22) at ($(H2)+(2\nd,-.4\nd)$)       {$L_{22}$};
   \path 
     (H2) edge [bend left] node {} (L21)
     (H2) edge [bend left] node {} (L22);
  
  \node[state]  (X1) at ($(H1)+(0,-1.4\nd)$)       {$X_1$};
   \path 
     (L11) edge [bend right] node {} (X1)
     (L12) edge   node {} (X1);
  \node[state]  (X2) at ($(H2)+(0,-1.4\nd)$)       {$X_2$};
   \path 
     (L21) edge [bend right] node {} (X2)
     (L22) edge   node {} (X2);

\gettikzxy{(O1)}{\ox}{\oy}
\gettikzxy{(X2)}{\xx}{\xy}
  \node[state,fill=webyellow]  (O2)   at ($(\ox,\xy)-(0,\nd)$)                 {$O_2$};
   \path 
     (X1) edge  node {} (O2)
     (X2) edge  [bend right]   node {} (O2);

  \node[state]  (plus) at ($(O2)+(\nd,-.2\nd)$)       {$+$};
  \node[state]  (minus) at ($(O2)+(2\nd,-.4\nd)$)       {$-$};
   \path 
     (O2) edge [bend left] node {} (plus)
     (O2) edge [bend left] node {} (minus);

\gettikzxy{(minus)}{\minx}{\miny}
  \node  (O3)   at ($(\ox,\miny)-(0,\nd)$)                 {$A,B$};
   \path 
     (plus) edge  node {} (O3)
     (minus) edge   node {} (O3);
     
  \node         (C1) [ left of=O1] {$Cycle\ 1$};
\gettikzxy{(C1)}{\cx}{\cy}
\gettikzxy{(X1)}{\xx}{\xy}
  \node         (C2) at ($(\cx,\xy)$) {$Cycle\ 2$};
\gettikzxy{(O2)}{\px}{\py}
  \node         (C3) at ($(\cx,\py)$) {$Cycle\ 3$};
\end{tikzpicture}
\\
\end{tabular}
}
{The theoretical parallelism (left) vs dynamic parallelism implemented on a processor system with runtime configurable  architecture (right).
}
{fig:flexibleproc}{}{}

\subsection{Dynamic parallelism}

Both utilizing a limited number of special multiple processing units,
and communicating through shared memory degrades the parallelism with respect to the theoretically reachable one.
Increasing the number of specialized processing units is possible, but 
(as Fig.~\ref{fig:SWvsHWparallelism} shows) in most of the general
purpose cases, those units cannot be fully utilized.
Communicating through a shared memory inserts new (obsolete) machine cycles with memory access,
and so (as Fig.~\ref{fig:dualproc} shows) the number of the cycles
needed for executing the task reduces disproportionally.
Both solutions strongly limit the available spedup, strongly increase the needed
resources and the dissipated power.
As an extreme case: the \gls{GPGPU}s outperform multiple-issue single processor
only 2.5 times, although about 100 times better performability is expected~\cite{Lee:GPUvsCPU2010}.
 
In the first case the cause of the inefficiency is the inflexible architecture,
in the second case the lack of any facility of intercore communication.
Let us suppose we have a kind of "on demand" type, flexible architecture, i.e. the processor
can provide the needed number of processing units for the operation, at the
expense of using some time for "renting" the needed unit(s).
The rented units are single-issue processors, but they are able to do both 
memory access and arithmetic operations. 
In Fig~\ref{fig:flexibleproc}
right side, it is assumed that the "cost of renting" is one fifth of a machine cycle.
At the very beginning the originating processor (in state $O_1$) notices that two multiplications
 shall be performed, so it rents \gls{PU}s $H_1$ and $H_2$, one by one (each in 0.2 machine cycle) for this goal.
 Those helper units notice they need two operands to load, so similarly they rent
 two more processing units only to do the  loading.

After loading, the helper processors receive their operands, so they can make the multiplication
in their second machine cycle and then deliver the result back  to the originating processor, which rents again
two more processors for the addition and subtraction, (in the third machine cycle
of the originating processor) for the last two operations. After the 3rd machine cycle,
both result operands are delivered back to the originating processing unit.
The execution time is longer than the theoretical 3 machine cycles.
In the figure  the total execution time is 3.8 machine cycles, and in the peak period,
6 simple-functionality single-issue processors are used.
\index{parallelism!dynamic}

Some quantitative parameters of the mentioned parallelization models
in the case of calculating our sample expressions are listed in Table~\ref{tab:ParallelParams}.
Since we have 8 operations, the single-thread execution time is 8. 
The average degree of parallelization is calculated as the ratio of 
the number of operations and number of cycles, and the efficiency is given as the 
ratio of speedup and number of \gls{PU}s.
As it can be expected, this dynamic parallelization model works in a way very similar
to the theoretical one, and its degree of parallelization
approaches the theoretical one (see Table~\ref{tab:ParallelParams}).

\MEtable{
\begin{tabular} 
{lrrrr} %
 \rowcolor{tableheadcolor}
 Parallelism model   & Parallelization & N      &Speedup      & Efficiency  \\
 \midrule
Theoretical          & $8/3=2.67$    & 4        & $8/3=2.67$    & $2.67/4=0.67$\\
Two-issue processor  & $8/7=1.14$    & 2        & $8/7=1.14$ & $1.14/2=0.57$ \\
Dual core            & $8/6=1.33$    & 2        & $8/6=1.33$    & $1.33/2=0.67$\\
Dynamic              & $8/3.8=2.11$  & $\approx 4.1$& $8/3.8=2.11$ & $2.11/4.1=0.51$ 
\end{tabular}
}{Parameters describing different parallelization models}{tab:ParallelParams}{}{}

\subsection{Requirements and consequences of dynamic parallelism}

The graph in Fig.~\ref{fig:flexibleproc} right side is essentially
the extension of the graph on the left side. 
The \gls{PU} is "rented" from some resource pool
and is returned after the operation finished, so in the next machine cycle it can be rented for a different goal.
It introduces some (trivial) dependence: before making an operation,
a processing unit must be rented; and after the operation finished,
it must be released, before starting the next cycle.
The renting process is transparent for the originating processing unit,
so the original dependence is preserved.
The states are in parent-child relationship: a parent can create
any number of children, but a child can have only one parent.
The parent remains responsible for performing the task it received, but it can delegate part of the task to its children.
If the parent  has delegated part of its job to the children,
it must wait until they terminate.

Since the operation is performed on a different \gls{PU}
rather than the original one, the complete state of the 
originating internals must be cloned into the created child unit
and after finishing the operation, part of the state (the result)
must be returned to the parent, in a synchronized way.

In the conventional architectures the machine cycles are uniform.
In the dynamic parallelization model the "all children ready" 
signal triggers the next cycle, which can be somewhat longer,
but can also be shorter, if in the child the last 
internal instruction stage is not utilized. 
With an effective pre-allocation mechanism, the time needed to 
allocate a helper core, can approach zero.

\MESourceFile[language={[y86]Assembler}, wide, options={numbers=left},
] {DynPar.Eyo}
{The dynamic parallelism implemented in assembly language for EMPA/Y86}{lst:DynPar}

\subsection{Mapping the operations to processing units}
The dynamic parallelism remains "theoretical" in the sense that
nothing limits the number of the needed processing units,
while in a physical system the number of \gls{PU}s is limited. 
The processing graph in Fig.~\ref{fig:DynPar8} exactly corresponds to the theoretical graph of dynamic parallelism in Fig.~\ref{fig:flexibleproc}, the 8th core cannot be utilized by the example code.
On a processor having finite number of \gls{PU}s the processing graph can be compressed horizontally,
at the price of increasing the number of the cycles, see 
Fig.~\ref{fig:DynPar4}.
When one keeps the dependence, some operations will simply
be postponed for a later machine cycle, prolonging the
processing time and decreasing the reached parallelism.
The \gls{PU}s are "rented" strictly for the time of performing
the processing step, so after a while "reprocessed" \gls{PU}s
get available. 
%

Obviously, the traditional fixed architectures are not able
to adapt themself to the task executed, so for that a special
architecture\cite{VeghEMPA:2016} must be used, which has some 
extra signals, storages and functionality,  see Fig.~\ref{fig:EMPAparentchild}. Such an architecture can be implemented
using methods known in reconfigurable technology,  like
using block RAMs, configurable wiring between fixed functionality blocks, etc.

Also, to program such a task special programming instructions are needed.
The code producing the processing diagrams (see section~\ref{sec:processingdiagram}) in Figs.~\ref{fig:DynPar8}-\ref{fig:DynPar4} is shown in Listing~\ref{lst:DynPar}.
Technically, one 'higher level' core is needed, which embeds the code
calculating the expressions in the sample.
The individual machine instructions are put in \gls{QT} frames~\cite{VeghEMPA:2016}
 only to provide complete visual analogy with Fig.~\ref{fig:flexibleproc}.
The same core could start reserving a helper core to load one operand; while waiting,
the core itself could load the another operand, and make the operation itself. 
The used method demonstrates, however,
that this kind of parallism can be extended towards both elementary operations
(like individual machine instructions) and complex expression evaluations 
(like the 8 elementary operations in the example code).
The operations in all cases can be independently executed,
and when discovering parallelism, no \gls{HW} limitations
shall be considered.

Here an event-controlled, rather than
clock-controlled, operation takes place, much similarly to the
pipelining and hyperthreading. This operating principle is not foreign from the
Neumann paradigms: there a new operation can only start when the
old operation frees the \gls{PU}.

\section{The tool chain for utilizing dynamic parallelism}
 
\subsection{The goal of the programming tools}
The \gls{EMPA} architecture not only provides a flexible \gls{HW}
for implementing dynamic parallelism (actually: provides an
end-user programmable architecture), but also provides other forms
of acceleration, like replacing certain machine instruction sequences
with using inter-core operating signals and replacing
(apparently non-parallelizable) operations with inter-core cooperation.
Those facilities are unusual in the conventional programming,
so a special programming approach had to be developed.
That approach must use conventional terms and programming interface,
to give chance to use higher-level languages for programming the 
\gls{EMPA} architecture, and at the same time must provide a way
to fully utilize the unconventional features of \gls{EMPA}.

\subsection{The Y86 processor}
\gls{EMPA} actually means some architectural principles, rather than a certain 
concrete processor or core. The work described here is based on using the Y86~\cite{hallaron}
processor as core. It is not a real processor in the sense that it has very few instructions
(finally, its purpose is educational). From our particular point of view it has advantages, like
\begin{itemize}
\item models a widely used architecture
\item it is simplified and without optimization
\item it has an open-source toolchain (simulator and assembler)
\item its \gls{ISA} allows to implement additional instructions and registers with easy
\end{itemize}

    As the first steps towards a tool chain, an \gls{EMPA}-aware ISA simulator and an assembler has been prepared~\cite{VeghEMPAthY86:2016}.
    These simple tools allow to prepare executable programs 
    for the \gls{EMPA} and characterize the performance features of the architecture~\cite{VeghEMPA:2016},
    as well as to develop and scrutinize further features.

\MEfigure[wide,resize]{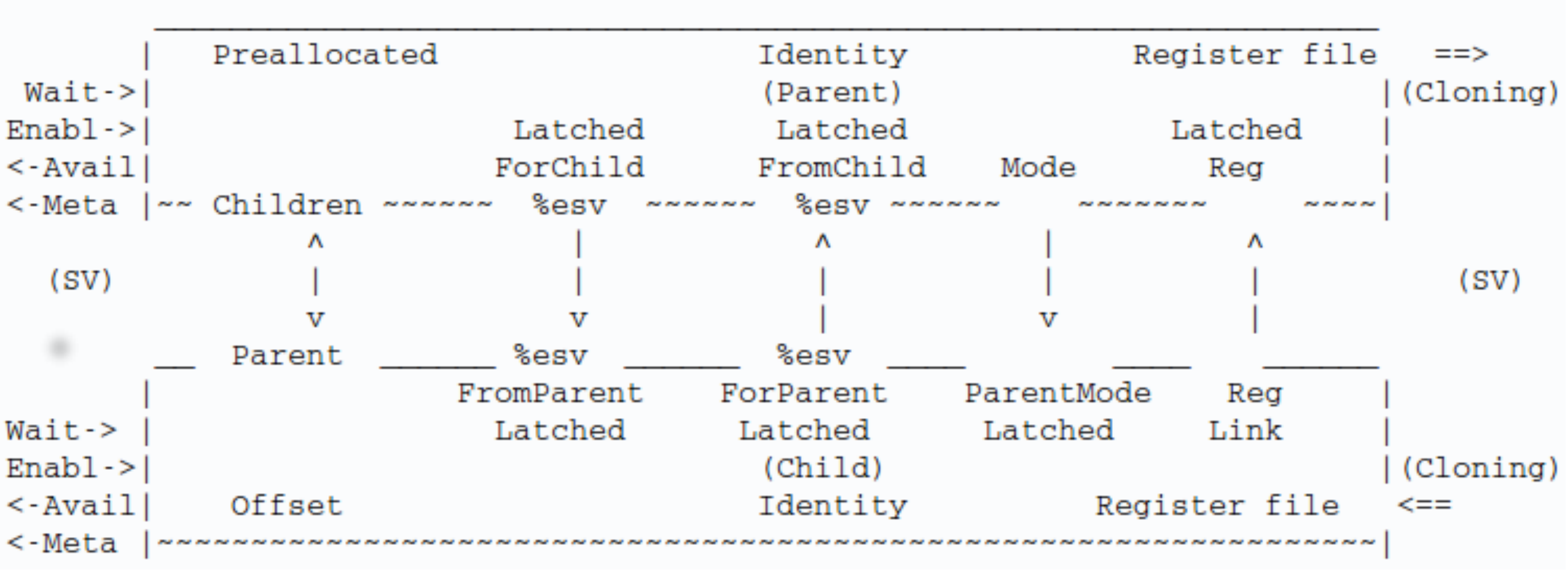}
{The communication signals and data between parent and child cores in EMPA}
{fig:EMPAparentchild}{}{}

For educational purposes, a simplified Intel X86 processor has been developed and made publicly available~\cite{hallaron}, including an \gls{ISA}-level simulator and an assembler.
Using a more avanced (non-educational) processor simulator would considerably 
extend the usability of the simulator.
However, those simulators are optimized for an absolutely different 
single-processor architecture, and also usually do not provide an easy path to
adding the needed extensions.
It is only fair to compare \gls{EMPA} to an unoptimized 
conventional processor. If the comparison is advantageous for \gls{EMPA},
similar, but different \gls{HW} accelerators will be developed
for this architecture, too, and those optimized architectures can again be fairly
compared to the optimized conventional architectures.

\subsection{Extensions to the Y86 ISA}

The original Y86 \gls{ISA}~\cite{hallaron} utilizes one-byte instruction codes, and in an unused instruction slot the group of metainstructions
\index{Y86}
utilized to configure the newly introduced  \gls{SV} control layer~\cite{VeghEMPA:2016}
 of the processor
has been implemented. Following the Y86 conventions,
a member of the \gls{EMPA} metainstruction group is coded as the group code in the high nibble and the 
member code in the low nibble. The mnemonic of a metainstruction always starts with a 'Q' (for  \gls{QT}).
The metainstructions can have zero, one, or two arguments, and their total
length is between one and six bytes. 

\subsection{The EMPA simulator(s)}

The simulator was written having electronic components
in mind, i.e. it operates in a cycle-accurate way.
The engine uses as core functionality the Y86 \gls{ISA} simulator,
 slightly extended.
 Both a command line based 
 and a Qt5~\cite{QTProgramming} based graphical interfaces  have been added to the simulator.
The GUI simulator is equipped with step-wise execution and logging; it produces processing diagrams like the one in Fig.~\ref{fig:EMPAdiagram},
and provides different kinds of statistics,
allowing to scrutinize the sophisticated
operation of the EMPA/Y86 processor, and to derive operational characteristics~\cite{VeghEMPA:2016}.

\section{Assembly extensions for EMPA}

The support for the unconventional \gls{EMPA} features is implemented
through a surprisingly low number of
 new assembly (meta)instructions and other extensions,
causing just a little difference relative to the traditional single-processor case.

\subsection{Creating and terminating quasi-threads}

The \gls{EMPA}-aware code is organized into special units of \gls{QT}s~\cite{VeghEMPA:2016},
of intermediate size and structure,
somewhere between the \gls{HW} unit 'machine instruction' and \gls{SW} unit 'thread'.
Handling \gls{QT}s  is supported 
by the new assembly instructions \lstinline|QCreate| and \lstinline|QTerm|,
which must be used in a bracket-like way.

The \lstinline|QCreate| instruction has two arguments.
The first one is the address of the matching \lstinline|QTerm|.
This informs the parent core, where to continue
after delegating the code in the \gls{QT}. The second argument is the \textit{link
register} (either a physical one or a pseudo-register, see section~\ref{sec:pseudoregisters}).

The \lstinline|QTerm| instruction has no arguments, but implies a \lstinline|QWait -1| and clones back the link register, defined by the matching \lstinline|QCreate|. 

Example (As shown, extensive labeling is utilized for referencing):\par
\noindent\lstinline|CLabel: QCreate TLabel, %eax|\par
\noindent\lstinline|... executable instructions ...|\par
\noindent\lstinline|TLabel: QTerm|\par

\subsection{Synchronizing QTs}

The assembler provides two kinds of instructions to support explicit waiting.
Instruction \lstinline|QWait| only considers its own children,
while \lstinline|QPWait| considers the sisters (the other children of the parent).
Upon finding a \lstinline|QxWait|, the requesting core gets blocked until
the referred to \gls{QT} terminates.
The wait instructions
have one argument. The argument is either the address of one 
particular \gls{QT} or a wildcard (All \gls{QT}s in the scope).

\MEtable{
\begin{tabular} 
{lll} %
 \rowcolor{tableheadcolor} Trigger    & Transfer in parent & Transfer in child  \\
 \midrule
QxCreate   &   ForChild$\rightarrow$ & $\rightarrow$FromParent  \\
   & register file$\rightarrow$ & $\rightarrow$register file \\
   & Mode$\rightarrow$ & $\rightarrow$ParentMode \\
QxWait (QTerm)  & link register$\leftarrow$  &    \\
QTerm & & $\leftarrow$link register\\
mass  processing & ForChild$\rightarrow$ & $\leftarrow$FromParent \\
(mode dependent)  & FromChild$\rightarrow$ & $\leftarrow$ForParent \\
\end{tabular}
}{The trigger signals and their effect on latched inter-core data communication}{tab:esvtriggers}{}{.9}

The \lstinline|QxWait| metainstructions play an important role also 
in synchronizing data transfer 
 between parent and child cores. 
As described in~\cite{VeghEMPA:2016}, the asynchronous operation of cores
needs well-designed transport policy, especially when using the link register.
Table~\ref{tab:esvtriggers} shows the triggered data exchange policy between the parent and child cores.

Example :\par
\noindent\lstinline| QWait  CLabel # Wait specific QT|\par
\noindent\lstinline| QPWait -1 #Wait all sister QTs|\par

\subsection{Subroutine call with EMPA}

The instruction \lstinline|QCall| provides a possibility to place 
the called \gls{QT} out of the body of the code.
The instruction has an address argument, where the called \gls{QT} begins.
The referred to \gls{QT} commences in the child core, the control in the parent
 returns immediately to the address next to \lstinline|QCall|. 
Since the argument of the metainstruction is the address  of a metainstruction
\lstinline|QCreate|, its functionality is automatically implied
in functionality of \lstinline|QCall|. Practically, the difference is
that the called QT code is located outside the body of the
parent control flow, resulting in a modular, clear program structure.

Notice that the return address, unlike in \gls{SPA} case, should \textit{not} be remembered:
the called subroutine runs on a new core and uses its own \gls{PC},
while the caller continues processing with instruction next to \lstinline|QCall|.
\textit{The \gls{HW} should not save the return address, in this way less memory cycles
and machine instructions needed, and also the \gls{HW} addresses are not interlaced
in the \gls{SW}-handled stack items.} This also reduces the need for calling frames
and simplifies addressing of automatic or passed variables.

Example:\par
\noindent\lstinline|QCall CLabel|\par

\subsection{Supporting cooperation between cores}\label{sec:cooperationEMPA}

Several classes of processor accelerators serve executing masses of
instructions in parallel. \gls{EMPA} provides mass processing modes for this goal.
Mass processing functionality is implemented through the metainstructions \lstinline|QAlloc|,
\lstinline|QTCreate| and  \lstinline|QFCreate|.

The two arguments of \lstinline|QAlloc| are a mode value, and a register,
containing the argument for the requested operation.
Since \lstinline|QAlloc| actually is a request to the \gls{SV} that the
requestor core wants to rent additional core(s), the program must prepare
for two answers, and handle them in two different branches.
The two branches are implemented by metainstructions   \lstinline|QTCreate|
and  \lstinline|QFCreate|. Exactly one of them will be executed
after the last \lstinline|QAlloc|, the other party will behave
as a NOP instruction (i.e. it follows an \lstinline|if..then..else| logic).

Both cases must be programmed:
there is one \gls{QT} prepared for the case when mass processing in the requested mode is possible,
and another one for the case when not.
These two metainstructions have the arguments
and functionality identical with those of  \lstinline|QCreate|,
except that they are only executed if the pre-allocation was successful
and was NOT successful, respectively.

Metacommand \lstinline|QAlloc| must also assure that the needed number of cores
will be available at some later time for the parent core. For this goal,
the needed cores are preallocated for the core: they will appear for the
concurrently working cores as unavailable cores, but  \lstinline|QTCreate|
can use them to start new \gls{QT}s.

If the core preallocation was successful (i.e. there are enough cores), the \lstinline|QTCreate|
branch will be executed as many times as needed.
This is controlled by the \gls{SV}, and is carried out in consecutive clock  cycles, one by one. The \gls{PC} of the parent will advance to the 
address next to the matching \lstinline|QTerm| only when  \lstinline|QTCreate| was processed the requested number of times,
in certain modes each time with a newly allocated  child core.
 In this way the \lstinline|QTCreate|  actually corresponds
to as many \lstinline|QCreate| metainstructions as many repetitions were requested 
by \lstinline|QAlloc|.
In these actions only the preallocated (rather than unallocated) cores are used. 

The functionality of \lstinline|QTCreate| is not simply making a replica of the parent
for the preallocated cores.
The pseudo-register \lstinline|%esv| behaves  as configured by the 'mode' argument of \lstinline|QAlloc|.
The \lstinline|QFCreate| is just a
wrapper for the instructions and metainstructions to be executed
when there are not enough cores for the given type of mass processing, i.e. the
processing must follow some other way. These conditional allocations can be nested.
Note that the link register for both branches must be the same.

Example:\par
\noindent\lstinline|        QAlloc mode, %edx|\par
\noindent\lstinline|Q4TC:	QTCreate Q4TT, %ecc|\par
\noindent\lstinline|        instructions if we have enough cores|\par
\noindent\lstinline|Q4TT:	QTerm|\par
\noindent\lstinline|Q4FC:	QFCreate Q4FT, %ecc|\par
\noindent\lstinline|		instructions if we do NOT have enough cores|\par
\noindent\lstinline|Q4TT:	QTerm|\par

\subsection{Pseudo-registers}\label{sec:pseudoregisters}
For implementing an effective data transfer between the cooperating cores,
 some pseudo-registers have been implemented.
 The pseudo-registers are seen by the \gls{ISA} as registers,
 but they represent not a simple storage. Rather,  
they might behave in an extraordinary way: they can transfer data 
synchronously between parent and child, in both directions, and they can 
change the data they provide for their partner between the consecutive accesses.
 
Register \lstinline|%eno| is used where the syntax requires the presence of a register argument,
but the related activity is not desirable.
\index{register!\%eno}
Register \lstinline|%ecc| is for returning condition codes only,
while \lstinline|%esv| is used for parent-child related activity.
\index{register!\%ecc}   
\index{register!\%esv}
While the first two pseudo-registers 
 can only be used as arguments of \lstinline|QCreate| (i.e. as link registers),
the functionality of \lstinline|%esv| largely depends
on the context it is used in, see Table~\ref{tab:esvmapping}.

\MESourceFile[language={[y86]Assembler},
options={lastline=38, numbers=left}
,wide,decorations={\MEListBalls{lst:Qasum04}{6,8,10,9,12,14,15}
}] {Qasum0_4.Eyo}
{The sum-up routine using NO EMPA facilities}{lst:Qasum04}

\MESourceFile[language={[y86]Assembler},
options={lastline=38, numbers=left}
,wide,decorations={\MEListBalls{lst:Qasum14}{4,7,8,9,10,13,11,12}
}] {Qasum1_4.Eyo}
{The sum-up routine using EMPA looping FOR facilities}{lst:Qasum14}

\MEtable{
\begin{tabular} 
{lll} %
 \rowcolor{tableheadcolor} Context    & As source & As destination    \\
 \midrule
Cloning (link register)  & ForParent & FromChild  \\
Child in mass processing   &   FromParent & ForParent  \\
Parent in mass PREprocessing   & FromParent & ForChild \\
Parent in mass POSTprocessing  & FromChild  &  ForParent   \\
Other (general) case & FromChild & ForParent\\
\end{tabular}
}{The context dependent mapping of register \lstinline|\%esv| to latched registers}{tab:esvmapping}{}{.9}

Fig.~\ref{fig:EMPAparentchild} (repeated from \cite{VeghEMPA:2016} for convenience) shows
how the parent and child cores communicate with each other using latch registers.
The \lstinline|%esv| register is mapped in a context-dependent way to the latched registers,
see Table \ref{tab:esvmapping}.
As shown in the Table, the mass processing parent role is divided into PRE-processing
(between \lstinline|QAlloc| and \lstinline|QTCreate|), and POST-processing
(between \lstinline|QTerm| of the child and \lstinline|QTerm| of the parent)
phases.
Using \lstinline|%esv| as link register,
the \gls{SV} reads the content of 'ForParent' in the child  and writes it
to 'FromChild' in the parent. This means, that if a child core wants to
transfer to its parent the data it received from its own child,
the child core must use an explicite \lstinline|rrmovl %esv, %esv| instruction.
Register \lstinline|%esv| is designed for helping cooperation,
and cannot be used as general purpose register.

\section{Algorithmic aspects}\label{sec:EMPAalgorithm}

It was early recognized~\cite{BackusNeumannProgrammingStyle}, that even our programming
languages are heavily influenced by the single-processor approach, and so are our algorithms.
The disclosed new possibilities in
the \gls{EMPA} architecture also need new thinking in designing the algorithms.
The synergy between the possibilities of \gls{EMPA}  and the new \gls{EMPA}-aware algorithms
(i.e. suggesting methods to implement in \gls{EMPA} which can simplify or boost 
old or develop more efficient new algorithms) can result in further performance
increase of our \gls{HW}/\gls{SW} systems.
\gls{EMPA} provides a couple of general frames and methods for using such possibilities,
as shown by the examples below,
and is ready to implement further such frames.  
Below, a simple programming example is presented in four different versions,
to illustrate how different accelerating principles~\cite{VeghEMPA:2016} of \gls{EMPA} can be 
used in practice.

\subsection{The conventional coding (or NO mode mass processing)}
The first mode is the NO mass processing mode.
It exactly matches the traditional programming:
NO real mass processing takes place,
no metainstructions are used and the required loop control
functionality is provided through calculations.
It requires only the original \gls{PU}, uses the 
same instructions and has the same execution time,
as the traditional programs.

In this code the operands are loaded immediately 
before the calculation. The summing is as simple as possible:
first the sum is cleared 
\ifx\Arxiv\undefined (\MEBall{Listing~\ref{lst:Qasum04}}{1}) 
\else (Listing~\ref{lst:Qasum04}, line 6)\fi,
 and the number of items verified
\ifx\Arxiv\undefined (\MEBall{Listing~\ref{lst:Qasum04}}{2}) 
\else (Listing~\ref{lst:Qasum04}, line 7)\fi.
These are one-time actions, not parallelized.

From beginning with "Loop", the usual activity takes place:
in addition to the payload operation
\ifx\Arxiv\undefined (\MEBall{Listing \ref{lst:Qasum04}}{3})%
\else (Listing~\ref{lst:Qasum04}, line 10)%
\fi\ \lstinline|addl %esi, %eax|,
using non-payload operations
the item is addressed, fetched~
\ifx\Arxiv\undefined (\MEBall{Listing \ref{lst:Qasum04}}{4})%
\else (Listing~\ref{lst:Qasum04}, line 9)%
\fi,
the address advanced to the  
next item
\ifx\Arxiv\undefined (\MEBall{Listing \ref{lst:Qasum04}}{5})%
\else (Listing~\ref{lst:Qasum04}, line 12)%
\fi,
the loop counter advanced and verified
\ifx\Arxiv\undefined (\MEBall{Listing \ref{lst:Qasum04}}{6})%
\else (Listing~\ref{lst:Qasum04}, line 14)%
\fi,
and  a conditional jump instruction
\ifx\Arxiv\undefined (\MEBall{Listing \ref{lst:Qasum04}}{7})%
\else (Listing~\ref{lst:Qasum04}, line 15)%
\fi\ 
closes the loop.
Upon exiting the loop, register \lstinline|%eax| contains
the sum 
\ifx\Arxiv\undefined (\MEBall{Listing \ref{lst:Qasum04}}{8})%
\else (Listing~\ref{lst:Qasum04}, line 16)%
\fi.

Notice that register  \lstinline|%eax| contains
the partial sum during the calculation.
This is the main source of inefficiency: in the 
payload operation \lstinline|addl %esi, %eax| the previous content of the
destination register is read, then the operation performed
and the result is written back as new content to the destination register.
Notice that the non-payload instructions have no role at all
for the calculation,
furthermore that we are interested in the final result only,
and not at all in the partial results.

\subsection{The basic loop: FOR mode mass processing}

As seen above, in such a simple loop the non-payload activities
require much more executable instructions, than the payload 
activity; and so: they take most of the processing time.
The overall performance can be enhanced through omitting
those service instructions as machine instructions,
and use \gls{HW}-implemented facilities instead.
\gls{EMPA} provides simple loop organization facility,
which helps to eliminate those non-payload instructions.

The first three machine instructions
\ifx\Arxiv\undefined \MEBall{Listing~\ref{lst:Qasum14}}{8}-\MEBall{Listing~\ref{lst:Qasum14}}{1}-\MEBall{}{2})%
\else (Listing~\ref{lst:Qasum14}, lines 4-7)%
\fi\ 
 are identical with those 
shown in listing \ref{lst:Qasum04}. The metainstruction 
\index{QAlloc}
\lstinline|QAlloc 1, %edx|  
\ifx\Arxiv\undefined (\MEBall{Listing \ref{lst:Qasum14}}{3})%
\else (Listing~\ref{lst:Qasum14}, line 8)%
\fi\  sets operating mode~\lstinline|1| (this is FOR),
preallocates one core
and tells \gls{SV} it wants to use the pre-allocated core \lstinline|%edx| (actually: 4) times for looping.
This core will be available for the requesting core only, and only
until looping is over.

The \gls{SV} clears in the parent the base index in 'ForChild' to be transferred to the rented child.
\index{register!ForChild}
This value will be incremented by 4 between the iterations,
so the actual child can always reach the actual offset value.
Since \lstinline|%ecx| contains the base address of the vector,
and the child inherits the register file of the parent,
the child could calculate the actual address from the
base and the offset. However, it takes time, so the pseudo-register
\lstinline|%esv| provides a useful facility to shorten the code.

In the pre-processing phase of the loop (between \lstinline|QAlloc|
and \lstinline|QTCreate|, see Table \ref{tab:esvmapping})
writing \lstinline|%esv|
\ifx\Arxiv\undefined (\MEBall{Listing \ref{lst:Qasum14}}{4})%
\else (Listing~\ref{lst:Qasum14}, line 9)%
\fi\ means
writing into 'ForChild'.
So, the instruction \lstinline|rrmovl %ecx, %esv| writes the base address into 
\index{register!ForChild}
'ForChild' in the parent. Since the contents of 
'ForChild' is increased by 4 between iterations,
the child receives a ready-made address, there is no need 
to make address calculation.

The metainstruction  \lstinline|QTCreate QT1LoopT, %eax| \ifx\Arxiv\undefined (\MEBall{Listing \ref{lst:Qasum14}}{5})%
\else (Listing~\ref{lst:Qasum14}, line 10)%
\fi\ will create child \gls{QT}. The \gls{SV} keeps the core pre-allocated  until loop terminates.
In this mode \gls{PC} of the parent core remains
pointing to \lstinline|QTCreate| 
\ifx\Arxiv\undefined (\MEBall{Listing \ref{lst:Qasum14}}{5})%
\else (Listing~\ref{lst:Qasum14}, line 10)%
\fi\ while the loop is running.
In the following clock periods,
the parent must wait, since the \gls{QT} running on the child core
is not yet terminated. When the child \gls{QT} finally terminates
and the \gls{SV} notifies the parent, it immediately executes
the next iteration, until the iteration count reached.

The payload activity is done by the child core,
i.e. by instructions between \lstinline|QTCreate| \ifx\Arxiv\undefined (\MEBall{Listing \ref{lst:Qasum14}}{5})%
\else (Listing~\ref{lst:Qasum14}, line 10)%
\fi\ and
\ifx\Arxiv\undefined (\MEBall{Listing \ref{lst:Qasum14}}{6})%
\else (Listing~\ref{lst:Qasum14}, line 13)%
\fi. Here the core fetches the actual argument
(\MEBall{Listing~\ref{lst:Qasum14}}{7})
from the address given by contents of its own pseudoregister
\lstinline|%esv|. In this mode reading \lstinline|%esv|
corresponds to reading 'FromParent' (see Table \ref{tab:esvmapping}),
which receives
\index{register!FromParent}
its contents from 'ForChild' in the parent
when \lstinline|QTCreate| is executed.

The child core inherits the internals (including contents of register file)
of its parent when the \gls{QT} is created, and returns the content of its link register
to the corresponding register in the parent when \lstinline|QTerm|
is executed, see Table \ref{tab:esvtriggers}. 
I.e. on entry (\MEBall{Listing~\ref{lst:Qasum14}}{8}) \lstinline|%eax| contains the
previous partial sum, on exit \lstinline|%eax| contains the
new partial sum, which will be cloned back to the parent,
and serves as the old partial sum in the next iteration.

Although not used in the present example, 
to provide a possibility to \lstinline|break| out of the loop,
the child can \textit{write} its own pseudo-register \lstinline|%esv|,
which means writing into 'ForParent'.
\index{register!ForParent}
The child can write \lstinline|0| into that register.
Upon executing \lstinline|QTerm|, the contents of that register 
will be written in register 'FromChild' in the parent.
Before executing \lstinline|QTCreate|, the \gls{SV} checks
 'FromChild' in the parent, and terminates loop if it is cleared,
\index{register!FromChild}
otherwise executes \lstinline|QTCreate| again: increases the address
in 'ForChild' and decreases count in 'FromChild'.
\index{register!FromChild}
\index{register!ForChild}
Of course, at the beginning \lstinline|QAlloc| writes the requested
number of repetitions into 'FromChild'.

Notice that the complete loop organization is accomplished by 
the \gls{SV}, on behalf of the parent core. The child's kernel can do any, much more complex activity. The only
limiting factor is that only the content of the link register
is back cloned to the parent.
Also notice that here actually no parallelization occurs. The parent is waiting
when the child is processing, and always only one child is used.
Another variant for FOR functionaly is to reserve a core for the individual 
vector elements. The child cores, as they would be created in adjacent cycles
in that mode, would receive the correct address from the parent. However, 
after termination they would overwrite \lstinline|%eax|, or would have to wait the termination of the previous \gls{QT} without performance gain.
Because of this, that mode cannot be used for summing up elements of a vector.
However, \gls{EMPA} has a more elegant and useful method for solving that problem.
 
  \MESourceFile[language={[y86]Assembler},options={numbers=left}, wide,
  decorations=
  {\MEListBalls{lst:Qasum54}{8,9,10,13,14,15}}
  ]{Qasum5_4.Eyo}
  {The sum-up routine using EMPA looping SUMUP facilities}{lst:Qasum54}

\subsection{The specialized loop: SUMUP mode mass processing}
 In summing up elements of a vector, 
the partial sum must be written back into a register in a
machine instruction, and read out the same again in the next cycle.
\textit{It is because the atomic unit in \gls{SPA} is the machine instruction.\textit{}}
Since for the time of looping a persistent connection can exist
between the parent and its children, \gls{EMPA} can provide
a way to eliminate this weakness, using its SUMUP mode.

The first three executed instructions are the same as in case
of Listing \ref{lst:Qasum04}.  The metainstruction \lstinline|QAlloc|
\ifx\Arxiv\undefined (\MEBall{Listing \ref{lst:Qasum54}}{1})%
\else (Listing~\ref{lst:Qasum54}, line 8)%
\fi\ now sets mode~\lstinline|5|, and \lstinline|%edx| now contains 
the number of requested helper cores (i.e this time we want to use several cores
in parallel, rather than one core several times). To spare time, the next
instruction
\ifx\Arxiv\undefined (\MEBall{Listing \ref{lst:Qasum54}}{2})%
\else (Listing~\ref{lst:Qasum54}, line 9)%
\fi\ overwrites the offset address
passed to the child
with the base address of the array, exactly, as in the case of FOR mode.
Also the same, that \gls{PC} in the parent will stay pointing to 
 \lstinline|QTCreate|
 \ifx\Arxiv\undefined (\MEBall{Listing \ref{lst:Qasum54}}{3})%
 \else (Listing~\ref{lst:Qasum04}, line 10)%
 \fi\ and creating children, one after the other.
It is, however, different, that several cores are preallocated at the beginning,
so the parent shall not wait: in the consecutive cycles
will create children, every time in a new core,
which child core will work in parallel with each other child cores  and the parent core.

\MESourceFile[language={[y86]Assembler},options={numbers=left}, wide,
decorations=
{\MEListBalls{lst:QasumC4}{8,10,16,21,25}}]{QasumC_4.Eyo}
{The sum-up routine using adaptive EMPA looping facilities}{lst:QasumC4}

The payload instructions (i.e. instructions between \lstinline|QTCreate| 
\ifx\Arxiv\undefined (\MEBall{Listing \ref{lst:Qasum54}}{3})%
\else (Listing~\ref{lst:Qasum54}, line 10)%
\fi\ and \lstinline|QTerm|
\ifx\Arxiv\undefined (\MEBall{Listing \ref{lst:Qasum54}}{4})%
\else (Listing~\ref{lst:Qasum54}, line 13)%
\fi\ are very similar to 
the case of FOR mode. The important difference is that now the
partial sum is 'stored' in register \lstinline|%esv|.
In this mass processing mode writing \lstinline|%esv| means
\textit{sending the data to a prepared adder in the parent}~\cite{VeghEMPA:2016}, where 
the addition is immediately executed, rather than reading the previous partial sum
from a temporary storage and writing it back. 
(Technically, it is written in 'ForParent' in the child,
but the instruction triggers copying the summand to 'FromChild'
in the parent, which is connected to one of the inputs of the adder,
while the other input is connected to the output of the adder.)
Both the old and new  partial sums are only
stored in the adder circuit, rather than in some special register.
The child \gls{QT}s are created with one clock cycle delay 
relative to each other, so they will send their fetched data
also with the same delay for the parent, so the adder can receive the data
and execute the addition without waiting or queuing.

The parent and its children will run in parallel, and after
starting the last child, the parent will continue with the 
instruction at address next to \lstinline|QTerm|
\ifx\Arxiv\undefined (\MEBall{Listing \ref{lst:Qasum54}}{5})%
\else (Listing~\ref{lst:Qasum54}, line 14)%
\fi. At that time
some of the children might still work, so here a \lstinline|QWait| metainstruction
must be used, otherwise the adder might contain not the final sum.
When all children terminated, the parent will be in post-processing phase,
so reading \lstinline|%esv| results in reading 'FromChild'
which latches the output of the adder. 
\index{register!FromChild}
The instruction \lstinline|rrmovl %esv, %eax| 
\ifx\Arxiv\undefined (\MEBall{Listing \ref{lst:Qasum54}}{6})%
\else (Listing~\ref{lst:Qasum54}, line 15)%
\fi\ will bring to light
the till invisible sum. Note, that in this mode the link register 
has no role, so \lstinline|%eno| is used in \gls{QT} creation
\ifx\Arxiv\undefined (\MEBall{Listing \ref{lst:Qasum54}}{3})%
\else (Listing~\ref{lst:Qasum54}, line 10)%
\fi.

\subsection{The adaptive processing}
When using \lstinline|QAlloc| 
\ifx\Arxiv\undefined (\MEBall{Listing \ref{lst:QasumC4}}{1})%
\else (Listing~\ref{lst:QasumC4}, line 8)%
\fi,
the successful execution is not granted at all.
The compiler cannot know in advance, how many cores will be available
when the metainstruction will be executed,  
from having as many cores as vector elements,
to having one core only, so it must prepare for all possible cases.
The metainstructions
\lstinline|QTCreate| and \lstinline|QFCreate| provide an
\lstinline|if...then...else| construction to solve this problem.
It means that the compiler must generate
code for all those cases, and the \gls{SV} chooses one according to 
the actual core availability.
As it will be shown in the example, these constructions can be nested.

This adaptive program is displayed in Listing \ref{lst:QasumC4}. Actually,
the kernels of the three previous programs are put together
into a special structure. 
The most performable
operating mode for summing up elements of a vector is SUMUP mode,
provided that there are enough cores available at the moment
when the summation must be executed. 
If the first \lstinline|QTCreate| 
\ifx\Arxiv\undefined (\MEBall{Listing \ref{lst:QasumC4}}{2})%
\else (Listing~\ref{lst:QasumC4}, line 10)%
\fi\ is not
successful, then a \lstinline|QFCreate| 
\ifx\Arxiv\undefined (\MEBall{Listing \ref{lst:QasumC4}}{3})%
\else (Listing~\ref{lst:QasumC4}, line 16)%
\fi\ is executed. Within this
latter block another  \lstinline|QTCreate|
\ifx\Arxiv\undefined (\MEBall{Listing \ref{lst:QasumC4}}{4}-\MEBall{Listing \ref{lst:QasumC4}}{5})%
\else (Listing~\ref{lst:QasumC4}, line 21-25)%
\fi\ \gls{QT} pair is located.
If less than 4 cores are available, then the program attempts to allocate
at least one more extra core 
(i.e. attempts to use the next, less performable, but also less resource-hungry operating mode).
If this also fails, then continues
with the conventional processing.

As shown, \textit{the compiler accounts for all possibilities,
and the \gls{SV} chooses the right code fragment}.
Anyhow: the program will run to completion, but its execution time
will depend on the actual availability of cores. 
This availability is not sensible for the external observer, 
only the different execution times (see Table~\ref{tab:effectivealphacores})
will be noticed.
I.e. the multicore \gls{EMPA} processor might appear as a super-power
single-core processor for the user.

\MEtable{
\begin{tabular}[resize]{rrrclll} %
 \rowcolor{tableheadcolor} Time	    & No of      & Speedup & $\alpha_{eff}$\\
 \rowcolor{tableheadcolor}  (clocks)	& cores (k)  &   (s)   & \\
 \midrule
 142  &   1  &  1    &     1 \\
 \midrule
156   &   2  & 0.91  &   -0.20 \\
156   &   3  & 0.91  &   0.65 \\
80   &   4  & 1.58  &   0.58 \\
 38  &   5+  &  3.74   &     0.92 \\
\end{tabular}
}{Speedup and effective parallelization for adaptive SUMUP (see Listing~\protect{\ref{lst:QasumC4}}), in function of the number of available cores}{tab:effectivealphacores}{}{.75}

In Table~\ref{tab:effectivealphacores}, the execution time of program
in Listing~\ref{lst:Qasum04} serves as the base of comparison.
The rest of the lines in the table show cases when during executing
the program given in Listing~\ref{lst:QasumC4} the processor has different number of available
cores.
The slight increase relative to row 1 in execution time is due to
executing the metainstructions: this is the price one has to pay for
running a program, designed for many-core systems, on a single-core system.
As shown in Table~\ref{tab:effectivealphacores}, in a system having 5 cores,
this summing is nearly 4 times quicker (as shown in Fig~\ref{fig:EMPAdiagram}, only a small fragment of the code is parallelized).
The column $\alpha_{eff}$~\cite{VeghAlphaEff:2016} also shows, that \gls{EMPA} is designed for
many-cores: the utilization efficiency increases with the increasing
number of cores used. A more detailed analyzis of performance of \gls{EMPA} is given in~\cite{VeghEMPA:2016}.

\MEfigure[resize]{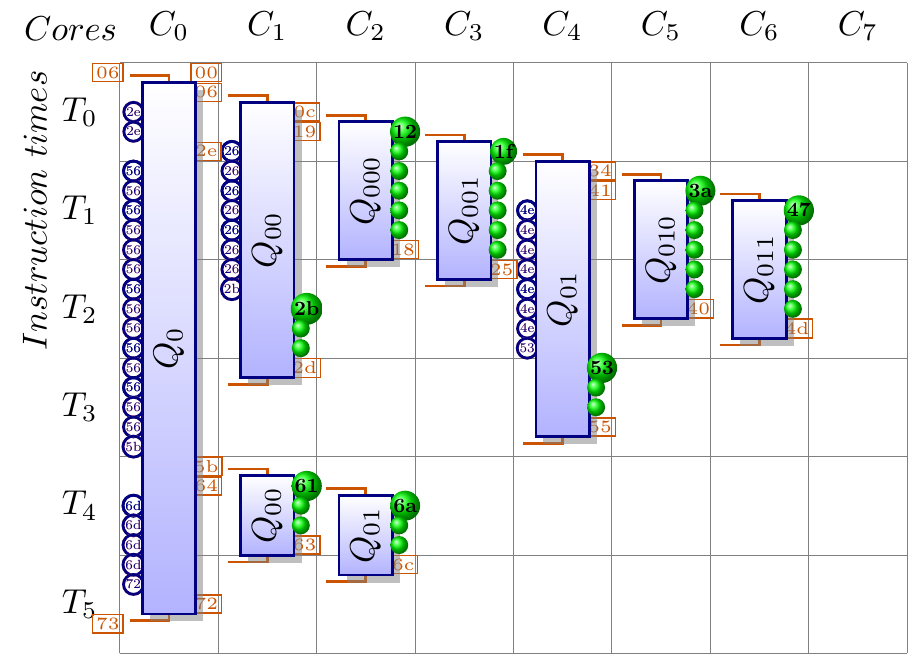}
{The processing diagram (compare to Fig.~\protect{\ref{fig:flexibleproc}}) of the sample program,  running on an 8-core EMPA processor}
{fig:DynPar8}{}{}

\section{The simulator}

The toolset has been published~\cite{VeghEMPAthY86:2016}
with online documentation. Because of the permanent development, it is still in alpha quality, but
it is usable. The unconventional features need careful utilization.

\subsection{The command line simulator}

The command line version of the simulator  runs to completion, and makes extensive logging in a file
showing all the details of the operation.
Although it is very useful during debugging,
and is inevitable when a power user attempts to "fine tune" his program,
for educational and demonstrational
purposes a  Qt5~\cite{QTProgramming} based graphical simulator has also been prepared. On the screen the complete internal life of the 
\gls{EMPA} architecture is displayed, as the cores are rented,
the intercommunication of the cores, etc.
The execution (in step-wise or run modes) can be followed.
%
%

\subsection{The processing diagram}\label{sec:processingdiagram}

The \emph{processing diagram} 
is a by-product of the simulators
and it attempts to visualize the rather sophisticated internal 
operation of the \gls{EMPA} processor.
The diagram should show, at which time, by which core, which instruction was executed; and how the cores interacted with each other; the cores can execute
conventional executable or metainstructions.
So, a lot of information must be crowded into the figure.

The diagram shows the cores on the horizontal axis
and the time on the vertical one. For better orientation,
grid lines are put to every 5th clock cycle.
The clock cycle here is the length of an \gls{SV} operation,
the instruction execution is supposed to be of variable length.
Arbitrary, but reasonable instruction lengths are assumed.

 The rectangular blocks represent
the \gls{QT}s, with hooks at the top and bottom, for their creation and termination.
In the columns $C_x$ the vertical rectangles represent the "lifetime" of a \gls{QT}. At the times outside the \gls{QT} rectangles,
the core is in power economy mode, not running a \gls{QT}.

The  parent-child relationship is illustrated with the labels of the \gls{QT}s: the first few chars are identical with those of the parent,
and the last char denotes the sequence number of the child.
On the figure (as well as in the simulator log files), for the human reader, core sequence numbers and textual \gls{QT} ID strings are shown
rather than the "one-hot" bitmasks used by the simulator.
 
\MEfigure[resize]{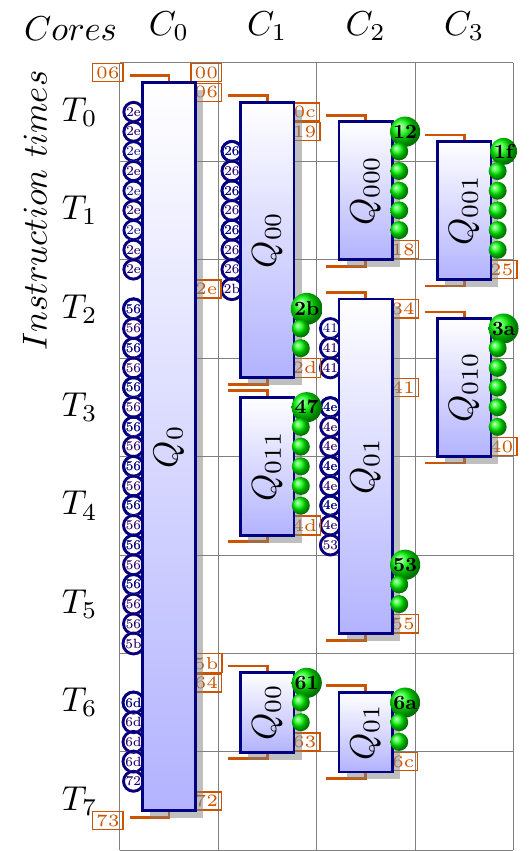}
{The processing diagram (compare to Figs.~\protect{\ref{fig:flexibleproc}} and \protect{\ref{fig:DynPar8}})) of the sample program, running on an 4-core EMPA processor}
{fig:DynPar4}{}{.60}

The memory address of a metainstruction is shown on the 
right side of the \gls{QT} in a rectangle,
the address of an executable command is shown on the top of a bigger ball, 
and some smaller balls represent the  duration of the instruction.
While a core is waiting, 
at the corresponding time a circle with the respective memory address is displayed at the left side of the \gls{QT} blocks.
From the memory address the source code can be found using the listings. 
Accessing pseudo register \lstinline|%esv| of parent by children is marked by an angular bracket,
 also showing the  direction of the transferred data.
 The places where summands are sent for their parent for summation, are marked 
 by an extra plus character.
%

\subsubsection{Dynamic parallelism}

The processing graph (see Fig.~\ref{fig:DynPar8}) is derived from the theoretical dependencies,
so the memory accesses within the cycles have no dependency on each other; i.e. the  memory can be accessed without limitation,
no need for assuring coherence, i.e. no need to use slow, power hungry and expensive shared memory. A memory of type like~\cite{Cypress15}
with several independent ports can solve the task.

The dynamic parallelism remains "theoretical" in the sense that
nothing limits the number of the needed processing units,
while in a physical system the number of \gls{PU}s is limited. 
The processing graph in Fig.~\ref{fig:DynPar8} exactly corresponds to the theoretical graph of dynamic parallelism in Fig.~\ref{fig:flexibleproc}, the 8th core cannot be utilized by the example code.

\MEfigure[]{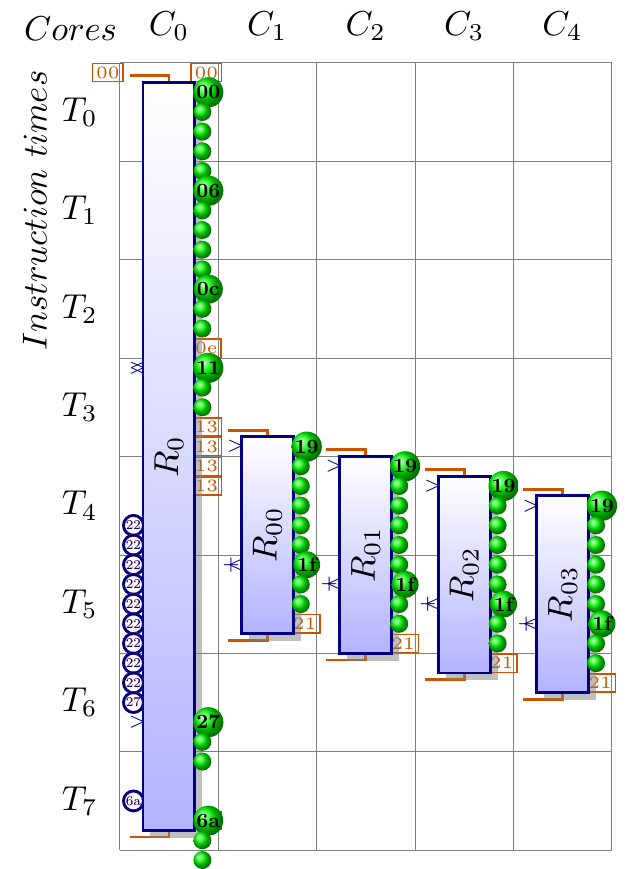}{Processing diagram of execution program in Listing \protect{\ref{lst:QasumC4}} running on 5 cores}{fig:EMPAdiagram}{}{.7}

On a processor having finite number of \gls{PU}s the processing graph can be compressed horizontally,
at the price of increasing the number of the cycles, see 
Fig.~\ref{fig:DynPar4}.
That diagram shows how the same code will be executed 
in a system with 4 \gls{PU}s only.
The first load operation can be started immediately, the second loading only when some \gls{PU} is released from the first operation.
The \gls{SV} in \gls{EMPA} notices that core $C_2$ finished the processing and is free again,
so the state $H_2$ is mapped to $C_2$ rather than to $C_4$ as in the case of 8 cores.  Anyhow, the operation
takes place when both operands are available. 

Some operations will simply
be postponed for a later machine cycle, prolonging the
processing time and decreasing the reached parallelism.
The programmer 
can give the theoretical 
dependency independently of the \gls{HW}, and the processing graph will adjust itself to the \gls{HW} at runtime.

\subsubsection{Vector sumup}
 Fig.~\ref{fig:EMPAdiagram} visualizes the  
operation of the \gls{EMPA} processor,
when running program shown in Listing~\ref{lst:Qasum54} on a 5-core system.
Notice how the inter-core operations (receiving the address
of the operand and sending the summand to the parent)
are shifted in time, and the final sum remains latent
until an explicit parent instruction takes it out into a
"visible" register.


\section{Conclusions} \label{sec:conclusions} 

The presented programming methodology demonstrates, how the 
dynamic parallelism and other unusual 
features of the \gls{EMPA} architecture can be supported in a programming style,
which is powerful enough to use the performance increasing facilities
of the architecture and at the same time remains close enough to the
conventional programming style.
This promises good hopes that the \gls{EMPA} architecture could be
effectively supported from high-level languages.

In the \gls{EMPA} implementation,
a special purpose "ad-hoc" computing unit is assembled on the fly, just for the
time of the (maybe complex) operation. This computing unit works with the 
maximum reachable parallelism, using the needed minimum of computing resources,
with maximum efficiency. 
The operation may be simple like loading an operand or
computing the two expressions used in discussing types of parallelisms,
or complex like summing up elements of a vector. 
The introduced extra signals and local storages allow to omit some instructions
used traditionally to organize a loop, and replace them with using internal signals.
The close vicinity of the \gls{PU}s (like in the case of modern many-core processors)
allows using cooperative regime for making calculations, and allows to parallelize even
the classic non-parallelizable sumup operation. This latter mode enables to gain
an order of magnitude in speedup. The programming facilities allow the programmers
(person or compiler) to use the unusual facilities offered by \gls{EMPA},
using  traditional terms and tools. Although the dependencies shall be
correctly considered, the hardware conditions should not be known at the time of coding:
the architecture follows the prescribed logic of programming, but adapts its resource need
to the momentary \gls{HW} availablility.

The real-time characteristics of processors also benefit from \gls{EMPA}.
To service an interrupt, no state saving and restoring is needed,
saving memory cycles and code. The program execution will be predictable: 
the processor need not be stolen from the running main process.
The atomic nature of executing \gls{QT}s will prevent issues like
priority inversion, eliminating the need for special protection protocols.

From the point of view of accelerators, an \gls{EMPA} processor 
provides a natural interface for linking special accelerators to the
processor. Any circuit, being able to handle data and signals shown in
Fig.~\ref{fig:EMPAparentchild} can be linked to an \gls{EMPA} processor with easy.

\bibliographystyle{IEEEtran}
\bibliography{IEEEabrv,Bibliography}

\begin{IEEEbiography}[{\includegraphics[width=1in,height=1.25in,clip,keepaspectratio]{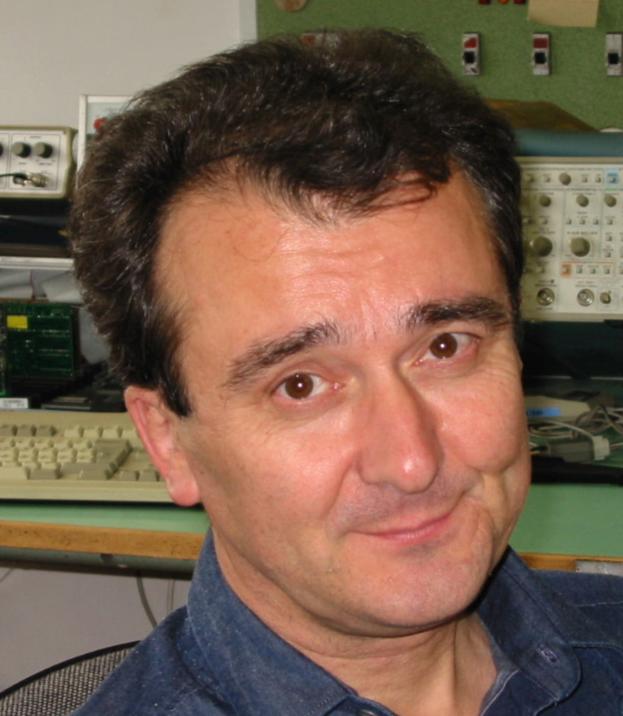}}]{V\'EGH, J\'anos}
received his PhD in Physics in 1991, ScD in 2006.
He is working in computing since the early 80's, and since 2008
is a full professor in Informatics, currently with University of Miskolc, Hungary.
In research, he is looking for ways to prepare more performant and 
more predictable computing, especially using mainly multi-core processors.
He is also dealing with soft processors and hardware-assisted operating 
systems, reconfigurable and many-processor computing. 
\end{IEEEbiography}




\end{document}